\documentclass[]{spie}  
\usepackage[]{graphicx}

\newcommand{\degree}{\ensuremath{^\circ}}

\usepackage{color}
\definecolor{blue}{rgb}{0.1,0.1,0.6}
\definecolor{orange}{rgb}{0.74,.35,0.099}
\definecolor{pale}{rgb}{0.90,0.90,0.95}
\definecolor{red}{rgb}{1.0,0.0,0.0}

\title{Gemini Planet Imager Observational Calibrations III: \\ Empirical Measurement Methods and Applications of High-Resolution Microlens PSFs }

\author{Patrick Ingraham\supit{ab}, Jean-Baptiste Ruffio\supit{cd}, Marshall D. Perrin\supit{e}, Schuyler G. Wolff\supit{e}, Zachary H. Draper\supit{f}, Jerome Maire\supit{g}, Franck Marchis\supit{c}, Vincent Fesquet\supit{h} 
\skiplinehalf
\supit{a}Kavli Institute for Particle Astrophysics and Cosmology, Stanford University, Stanford, CA 94305, USA \\
\supit{b}Department de Physique, Universit\'{e} de Montr{\'e}al, Montr\'eal QC H3C 3J7, Canada \\
\supit{c}SETI Institute, Carl Sagan Center, 189 Bernardo Avenue, Mountain View, CA 94043, USA\\
\supit{d}Institute Superieur de l'Aeronautique et de l'Espace, Toulouse, France \\  
\supit{e}Space Telescope Science Institute, 3700 San Martin Drive, Baltimore MD 21218 USA \\
\supit{f}University of Victoria, 3800 Finnerty Rd, Victoria, BC, V8P 5C2, Canada \\
\supit{g}Dunlap Institute for Astrophysics, University of Toronto, 50 St. George St, Toronto ON M5S 3H4, Canada\\
\supit{h}Gemini Observatory, Casilla 603, La Serena, Chile \\
}
\authorinfo{Further author information: contact Patrick Ingraham E-mail: patricki@stanford.edu }

\begin{document}
\maketitle 

\begin{abstract}
The newly commissioned Gemini Planet Imager (GPI) combines extreme adaptive optics, an advanced coronagraph, precision wavefront control and a lenslet-based integral field spectrograph (IFS) to measure the spectra of young extrasolar giant planets between 0.9-2.5 $\mu$m. Each GPI detector image, when in spectral model, consists of $\sim$37,000 microspectra which are under or critically sampled in the spatial direction. This paper demonstrates how to obtain high-resolution microlens PSFs and discusses their use in enhancing the wavelength calibration, flexure compensation and spectral extraction. This method is generally applicable to any lenslet-based integral field spectrograph including proposed future instrument concepts for space missions.
\end{abstract}


\section{Introduction}
\label{sec:intro}

The Gemini Planet Imager (GPI)\cite{Macintosh14a}\cite{Macintoshthis} is a newly commissioned instrument that combines an extreme adaptive optics system, an advanced coronagraph, precision wavefront sensing and a lenslet-based integral field spectrograph (IFS\cite{Chilcote12a}\cite{Larkinthis}) to measure the spectra between 0.9-2.5 $\mu$m of young extrasolar giant planets orbiting relatively nearby stars. Each GPI image consists of $\sim$37,000 microspectra, simultaneously imaged on the detector, each requiring individual calibration prior and extraction. The shape of these microspectra are defined not only by their input intensity, but also the point-spread function (PSF) of each individual microlens, combined with their location on the detector. Knowledge of the microlens PSFs is important for many aspects of the data reduction, including wavelength calibration, spectral extraction, characterization of crosstalk and understanding our spectral resolving power. 

Having a detailed model of the microlens PSFs enables two key applications: a robust spectral extraction and a precise wavelength calibration. For example, in order to create data cubes of sufficient quality to ensure an average uncertainty spectral characterization of 5\% (one of the primary GPI science goals), the error in the wavelength solution must be less than 1\%. The wavelength solution is determined from measuring the position of spectral lines of know wavelength using arc lamps. Although fitting these peaks using a Gaussian is sufficient to obtain the 1\% accuracy, this error can be greatly reduced if the centroiding method is improved to utilize the true lenslet PSF, as imaged on the detector. The reason for this improvement is the ability to eliminate the effect of pixel-phase error, a systematic positioning error that is a function of the sub-pixel position of the intensity peak relative to the pixel center. Pixel-phase error primarily affects under-stamped data, but is often also measurable in Nyquist sampled data.

The key challenge in obtaining a high-precision wavelength calibration with GPI is performing accurate centroiding of the spectral peaks of the arclamp calibrations because the microlens PSFs are either under- or critically sampled. Their positions are further complicated by the intrapixel sensitivity of the HAWAII-2RG detectors, which has been observed to decrease by 10-15\% toward the pixel edges\cite{Hardy08a}. Accurate centroiding of under-sampled images requires a high-resolution model that includes the instrumental microlens PSF combined with a knowledge of the detector properties. Previous works independently derive or assume these model components \cite{Zimmerman11a}, whereas the method we use here directly measures the combination of the two. This model, called the high-resolution effective PSF (ePSF), is constructed using a technique was developed by Anderson and King (2000) to fit the under-sampled PSFs of the Wide Field Planetary Camera 2 imager aboard the Hubble Space Telescope \cite{Anderson00a}. We have adopted this technique and modified it to permit the high-resolution reconstruction of our under-sampled microlens PSFs.

In this paper, we demonstrate a method to construct high-resolution narrowband microlens PSFs for GPI by combining sub-pixel dithers that are caused by internal flexure of the GPI IFS. The integration of the high-resolution PSFs into the GPI Data Reduction pipeline are then discussed, along with their applications in improving and or correcting multiple systematic errors introduced as part of the datacube reconstruction. We then discuss how this technique to determine the microlens PSFs can be used on current and future instruments and how to employ this technique to determine the microlens PSFs for current and future ground- or space-based instruments that include a lenslet-based IFS.

\section{Outline of GPI}

The GPI instrument houses a cryogenic ($\sim$80 K) IFS that contains both a dispersive prism for spectroscopic observations and a Wollaston prism for polarimetric observations. GPI's rectangular 2.8$\times$2.8 arcsecond field of view is divided into $\sim$37,000 sections (192$\times$192) using a transmissive microlens array with each $F$/200 lenslet having a 110 $\mu$m pitch. Each microlens forms an unresolved image of the pupil. These images are subsequently re-imaged and passed through either the Wollaston prism or the spectral prism before being re-imaged onto the detector. Due to space constraints, the re-imaging optics are highly powered and therefore the resulting field dependent aberrations elongate and or deforms the individual microlens PSFs as one moves away from the optical axis. Naturally, minute fabrication errors in the lenslet array cause small variations between the individual microlens PSFs, however, the distortion terms from the re-imaging optics is the effect and is primarily responsible for the variations measured on the detector. Figure 1 illustrates how the PSFs change as a function of field position. The PSFs in the figure are the derived high-resolution ePSFs rather than the detector sampled PSFs, this is done for clarity in demonstrating the effects of the field dependent aberrations. How these ePSFs are derived is discussed further in \S \ref{sec:construction}.

\begin{figure}[ht]
\begin{center}
\includegraphics[height=7.9cm, trim=0cm 0cm 0cm 0cm]{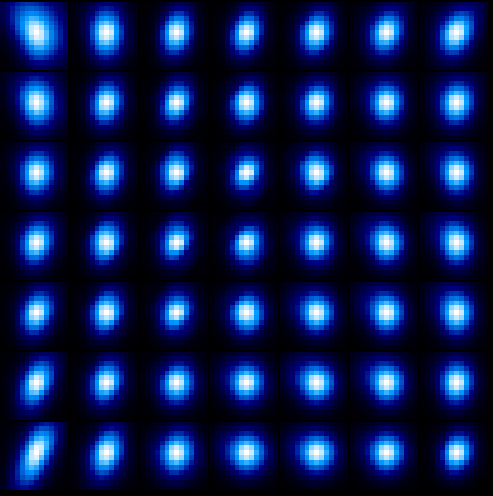} 
\end{center}
\caption[example]{ \label{fig:PSF_fov} The variation of the $Y$-band (1.000 $\mu$m) microlens PSFs across the field of view. This image shows the derived high-resolution models rather than the detector sampled images for clarity. The field dependent aberrations dominate the PSF shape towards the edges of the field.}
\end{figure} 

The microlens array has only a few of lenslets with either very low throughput or large distortions, for the purposes of this paper they have been masked, and are unused. In spectral mode, the microspectra are separated by 4.5 pixels in the spatial direction and 23.5 pixels in the dispersion direction. In polarization mode, the microlens PSFs are separated by $\sim$7.2 pixels \cite{Chilcote12a}.

\section{Flexure of the IFS}

The Gemini telescopes utilize an altitude-azimuth mount and Cassegrain design and therefore the instruments are mounted behind the primary mirror and move with the telescope during observations. Under normal operations, the instruments both move in elevation to track the target and rotate in position angle to follow sidereal rotation. With GPI, the instrument only exhibits motion in elevation because we always employ the Angular Differential Imaging technique, where the field rotator is disabled and the instrument is held at a constant position angle\cite{Marois06a}. As expected, the instrument exhibits small amounts of flexure due to the changing gravity vector. Prior to entering the IFS, the pointing and centering of the beam is corrected for flexure effects using look-up tables and two pairs of pointing and centering mirrors. Without compensation, this flexure effect would result in the star falling in a different position in the focal plane, and therefore on different lenslets. Once the beam propagates through the lenslet array, there is no further motion compensation. Therefore, as the instrument is moved into different positions, minute flexure motions of the optical elements result in a small shifts in the positions of the microspectra on the detector. The effect of flexure inside the IFS was predicted during the instrument design phase and therefore significant effort was made to ensure the IFS optical bench was stiff and resistant to flexure effects. 

The motion of the microspectra as a function of telescope elevation was measured by mounting the instrument on the flexure rig at the Gemini Telescope. The motions were measured to follow a semi-repeatable curve, with motion in the x-axis (spatial direction) of $\sim$1 pixel and motion in the y-axis (dispersion direction) of $\sim$0.5 pixels. A non-repeatable component was also observed when the instrument was moved into the nearly vertical position (elevation$<$5\degree), or when the instrument was rotated using the Cassegrain rotator. Although these instrument positions will not be encountered during GPI observations, they will occur during observations with other instruments, when the telescope is parked during the day, or during slews between targets. 

\begin{figure}[ht]
\begin{center}
\includegraphics[height=7.9cm, trim=0cm 0cm 0cm 0cm]{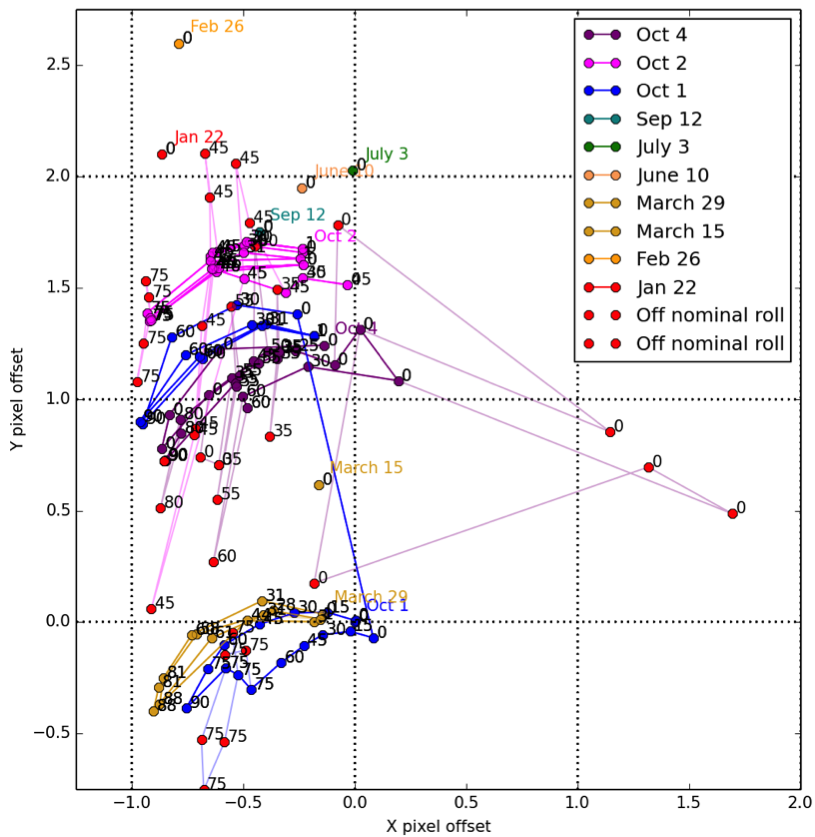} 
\end{center}
\caption[example]{ \label{fig:flexure} The relative position of the $H$-band microspectra as a function of instrument position and time. Between cool-downs, the spectra fall at different positions, but then follow a repeatable curve when moved in elevation (shown as the number beside each dot). If the Cassegrain rotator is moved, resulting in an off-nominal roll position (red dots), a non-repeatable offset is observed. The flexure motions enable a sub-sampling of the microlens PSFs.}
\end{figure} 

\begin{figure}[ht]
\begin{center}
\begin{tabular}{cc}
\includegraphics[height=7.9cm, trim=0cm 0cm 0cm 0cm]{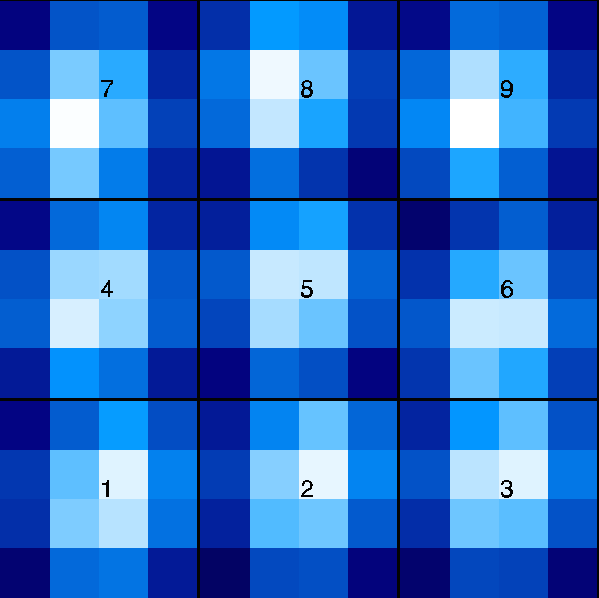} & \includegraphics[height=7.9cm, trim=0cm 0cm 0cm 0cm]{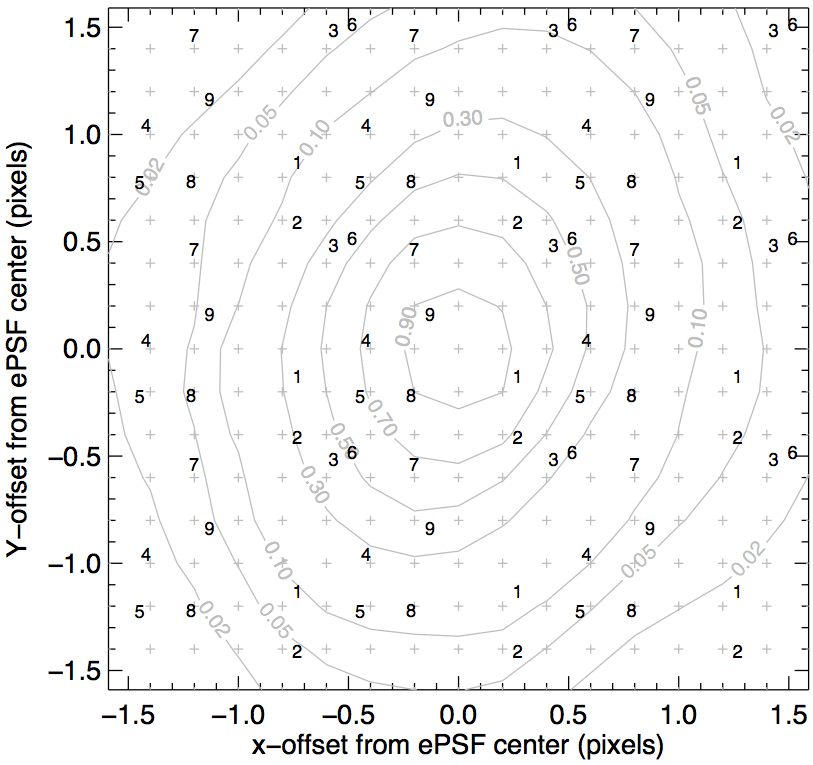}
\end{tabular}
\end{center}
\caption[example]{ \label{fig:pixel_sampling} \textbf{Left:} The same microlens PSF as seen on the detector, sampled differently due to 9 different instrument positions and the resulting internal flexure of the IFS. \textbf{Right:} A demonstration of how each microlens samples different parts of the high-resolution ePSF model. Each number designates the sample point from the detector PSF on the left image. The contours show the intensity levels of the high-resolution ePSf model, constructed using $\sim$1,200 detector PSF samplings.}
\end{figure}

Figure \ref{fig:flexure} shows the positions of the microspectra for multiple dates and multiple instrument elevations. All points of the same color indicate different installations on the flexure rig or different instrument supports, therefore large positioning changes of the instrument. For the majority of the cases, the flexure follows the repeatable curve. Positions of off-nominal roll, which indicate movement of the Cassegrain rotator, are shown as red circles. Information on how our observing strategy and wavelength calibration compensates for these effects can be found in Wolf et al (this conference\cite{Wolffthis}), however, what is important, is that this flexure effect results in the microspectra moving by sub-pixel amounts on the detector. This is analogous to observing a field of stars, but having a slight dither of telescope between observations, as discussed in Anderson et al (2000) \cite{Anderson00a}. Therefore, the same star (in our case microlens PSF) exhibits a slightly different sampling. Figure \ref{fig:pixel_sampling} shows a narrowband observation the same microlens PSF for multiple instrument positions. The different samplings of the same PSF allow us to reconstruct a high-resolution microlens ePSF.

\section{Construction of the High-Resolution ePSF}
\label{sec:construction}

To derive a high-resolution ePSF we employ a nearly direct implementation of the algorithm described by Anderson et al (2001). For their application, Anderson et al. derive high-resolution ePSFs from the undersampled data of the Hubble Space Telescope's Wide Field Planetary Camera 2 (WFPC2), then use them to perform precise astrometric measurements of stars. As mentioned in the introduction, this method of constructing a model is important to relieve the position error caused by the undersampling, which is found to be systematic in nature. Because the microlens PSFs of GPI are also under- or critically sampled, depending upon the wavelength, these systematic errors are also present in our data. Although GPI is very different from WFPC2, the same algorithm is applicable with only slight modifications. 

This paper does not describe the details of the algorithm, only it's implementation and the slight differences required to apply it to GPI data. In short, the Anderson et al algorithm utilizes the fact that a given pixel samples a single point of the ePSF (the effective PSF (ePSF)) which includes both the instrumental PSF, defined by the optical system, and the properties of the detector (e.g. pixel size and response). For WFPC2, each different sampling of a star in the same region of the science detector provides a single data point for the construction of a high-resolution ePSF. Several more samplings are obtained from images of same field but at multiple dither positions. The high-resolution ePSF is formed by averaging these samplings onto a user-designated grid. The high-resolution ePSF is then used to determine new, higher-precision positions of the stars in every image. The process is iterated until the systematic position errors of the PSFs on the detector, known as the pixel phase error, is eliminated. The pixel phase is defined as the position of the PSF peak relative to the center of the pixel. Pixel phase error refers to a systematic positioning error that occurs depending on where the peak falls relative to the center of the pixel. The reader is advised to consult the Anderson et al article for a detailed explanation of the process and systematic error.

With GPI, the PSFs formed by the individual lenslets in a single image are analogous to imaging multiple stars, whereas the images obtained in multiple instrument positions with different positional offsets due to instrument flexure, is analogous to multiple telescope pointings. By combining the two datasets, according to the methods derived by Anderson et al (2001), we derive a grid of 25$\times$25 high-resolution ePSFs over the field of view for each waveband. This sampling is sufficent to capture the variation of the microlens PSFs over the field and maintains a high number for samples in deriving each ePSF. The final narrowband images used to derive the ePSFs consisted of multiple ($>$5)detector images, each consisting of 5 coadds, at several instrument positions. The number of images was dictated by the intensity of the PSFs and the exposure time, but sufficient data was taken to ensure an SNR of 100 was obtained for all pixels within a 2 pixel radius of the peak. The number of instrument positions varied depending upon the band, ranging from 18-30. Both variation in elevation and Cassegrain rotator positions were used.

When constructing the high-resolution ePSFs, each detector pixel was sub-sampled into a 5$\times$5 grid. The high-resolution ePSFs are smoothed using a 3$\times$3 kernel analogous to the one shown in Anderson et al (2001), but modeled as a Moffat profile with a steep exponential. This smoothing is performed to minimize the effects of non-uniform sampling around a given ePSF gridpoint and to ensure the ePSF is a smoothly varying function. An important characteristic when considering the construction the high-resolution ePSFs was the 4.5 pixel spacing of the microlens images. This meant that every 2nd PSF is nearly identical, which is not ideal for sampling the PSFs. For this reason, only a 7$\times$7 array of lenslet PSFs, centered on the microlens of interest, were used in creating each high-resolution ePSF. Furthermore, using a box size $\sim$2$\times$ larger than this begins to show differences between the individual ePSFs due to the field dependent aberrations.

Creating a high-resolution ePSF from multiple flexure positions necessitates a transformation between frames. For the case of extracting spectra using our current extraction algorithm, a simple global x-y offset is sufficient. For our more advanced extraction algorithms\cite{Zackthis}, our wavelength determination\cite{Wolffthis}, and the creation of the high-resolution ePSFs, higher-order transformation terms are necessary. The transformations are determined by fitting each detector PSF in every flexure position with the high-resolution PSF, then using a box-smoothing algorithm to average the positional ePSF offsets between each flexure position and the reference position. During the iterative process, both the high-resolution ePSFs are improved, as well as the transformations. To date, 5 iterations have been sufficient to reduce the systematics in the PSF positioning to negligible levels. In the future, the higher-order flexure transformation frames will most likely be included in the spectral extraction algorithms.

\section{Examination of the ePSF}

\begin{figure}[ht]
\begin{center}
\includegraphics[height=5.5cm, trim=0cm 0cm 0cm 0cm]{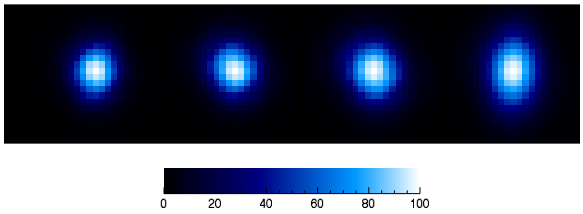} 
\end{center}
\caption[example]{ \label{fig:highres_psfs} A single high-resolution ePSF model derived at different wavelenghts (from left to right, $Y$, $J$, $H$, $K1$). These ePSFs are for a lenslet near the center of the array, where the field dependent aberrations from the camera optics are minimal and the chromatic terms from the microlens dominate the ePSF variation.}
\end{figure}

The microlens PSFs change not just over the field of view, but also with wavelength, as shown in Figure \ref{fig:highres_psfs}. The lenslet chosen for this figure is near the center of the array where the field dependent aberrations are minimal. With the exception of the $K1$ band, each ePSF (and detector PSF) is moderately square, as is expected due to lenslet array being composed of square lenslets. In the case of $K1$ band, the spectral bandpass of the 1\% narrowband filter is actually resolved, which creates the elongation in the spectral dispersion direction (vertical). Under normal GPI operation, the detector sampling limits the spectral resolving power ($\mathcal{R}$) in $K1$ band to be $\mathcal{R}$$\sim$70, however, with the high-resolution ePSF, this limitation is removed and the spectral resolving power is limited by the lenslet PSFs. Although this effect is an excellent demonstration of the ability to create high-resolution ePSFs from sub-sampled data and increase spectral resolving power, the fact that the line is resolved means that our high-resolution microlens PSF model in $K1$ band is not purely diffraction limited. Using a laser source would resolve this issue.

\begin{figure}[ht]
\begin{center}
\includegraphics[height=6.5cm, trim=0cm 0cm 0cm 0cm]{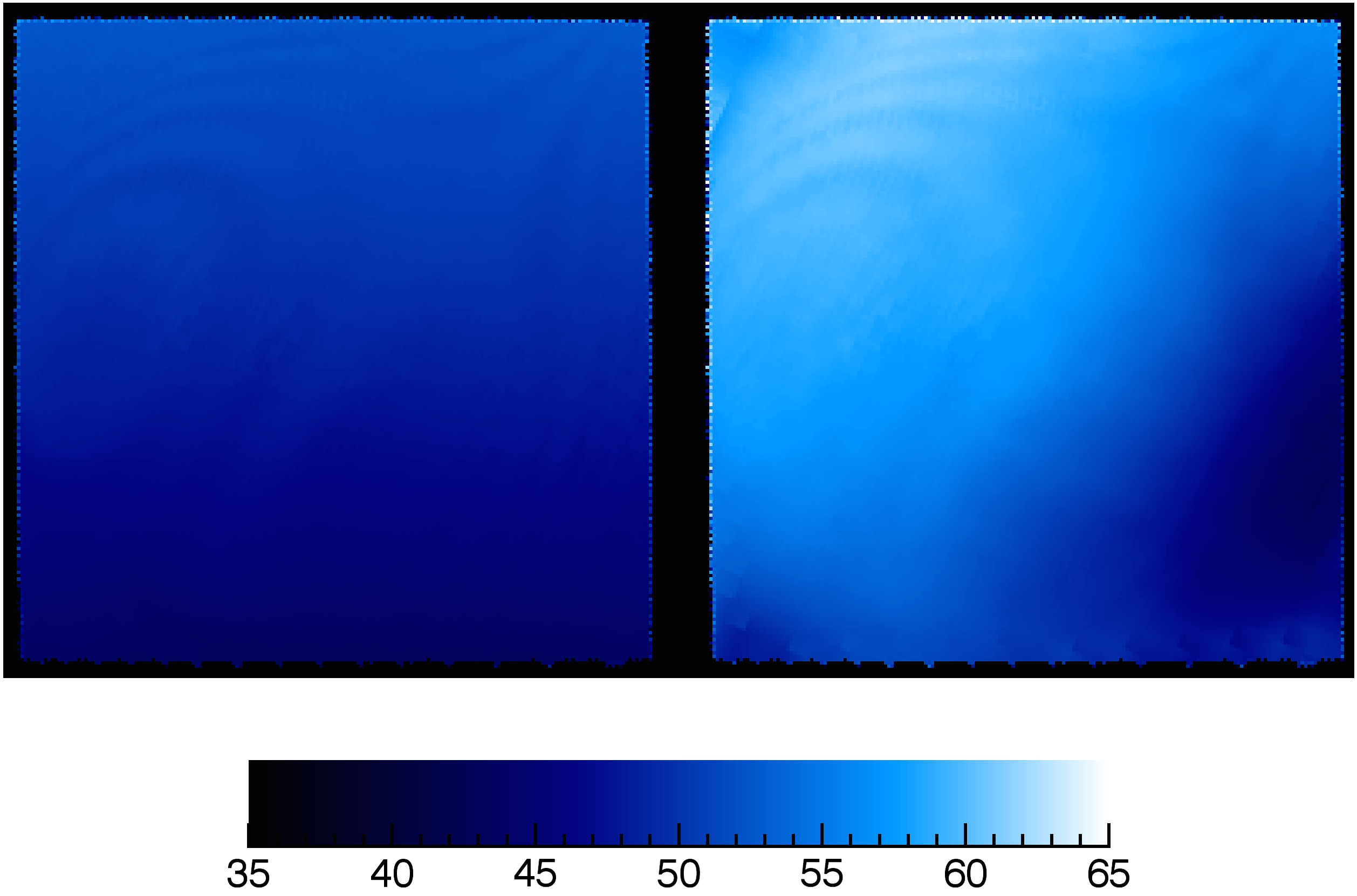} 
\end{center}
\caption[example]{ \label{fig:resolution} \textbf{Left:} The GPI spectral resolution over the field of view is defined by the length of the microspectra and the detector sampling (42$<$$\mathcal{R}$$<$53 in $H$ band). \textbf{Right:} The spectral resolution if the spectral resolution were to be limited by the microlens PSF (43$<$$\mathcal{R}$$<$62 in $H$ band).}
\end{figure}

The form of the PSF and how it evolves with wavelength is highly dependent upon its position in the field. Figure \ref{fig:highres_psfs} shows the ePSF as a function of wavelength for a lenslet located near the center of the field of view where the field dependent aberrations are minimal and thus ePSF evolves radially with wavelength. As one moves towards the edges of the field, the ePSF shape is dominated by the field dependent aberrations, as shown in figure \ref{fig:PSF_fov}. The effect is the most apparent in $Y$ band and decreases at longer wavelengths due to the decreased phase errors as a result of the optical path differences remaining constant. Due to the GPI optical design, the spectra are slightly shorter in the bottom of the field than the top, which when combined with our detector sampling, results in variation in our spectral resolution (42$<$$\mathcal{R}$$<$53 in $H$ band). However, if our spectral resolving power were to be limited by the size of the microlens PSFs, this would result in a higher spectral resolving power and a stronger variation over the field of view (43$<$$\mathcal{R}$$<$62 in $H$ band), as shown in Figure \ref{fig:resolution}. 

It is conceivable to develop a spectral extraction technique whose spectral resolving power surpasses the limit imposed by the detector pixels and is limited by the lenslet PSFs. The full width at half maximum (FWHM) of the GPI PSF on the microlens array spans $\sim$3-4 lenslets and the the adjacent microspectra are offset by 4.5 pixels. Therefore since the microspectra within a given FWHM of a star should be identical but sampled differently, you can construct a higher-resolution spectrum. Whether or not this is practical scientifically is highly dependent upon the signal to noise of the companion and if the increased spectral resolution would result in a significant gain in characterization ability.

\begin{figure}[ht]
\begin{center}
\begin{tabular}{cc}
\includegraphics[height=4.5cm, trim=0cm 0cm 0cm 0cm]{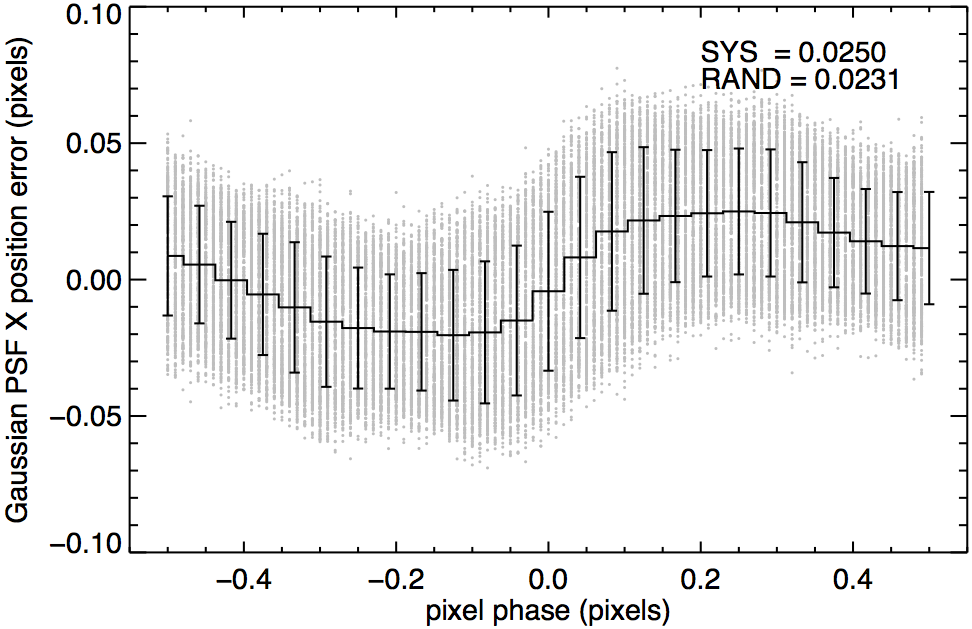} & \includegraphics[height=4.5cm, trim=0cm 0cm 0cm 0cm]{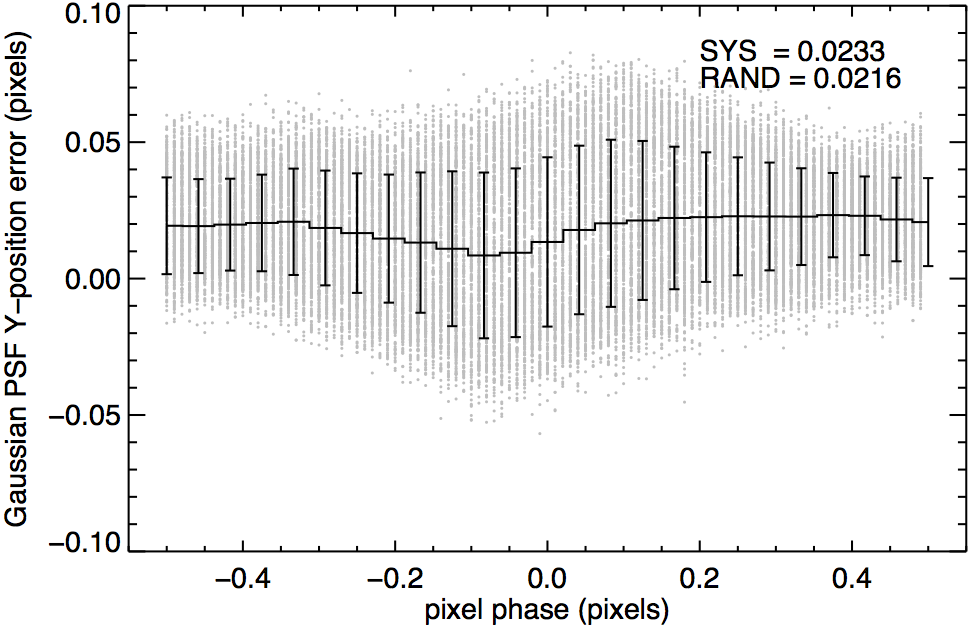} \\
\includegraphics[height=4.5cm, trim=0cm 0cm 0cm 0cm]{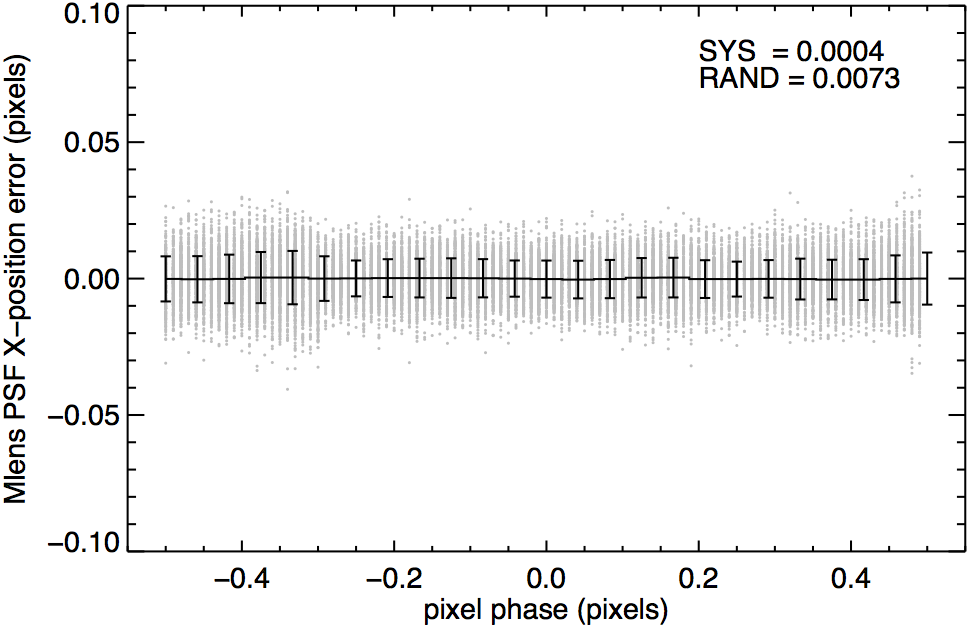} & \includegraphics[height=4.5cm, trim=0cm 0cm 0cm 0cm]{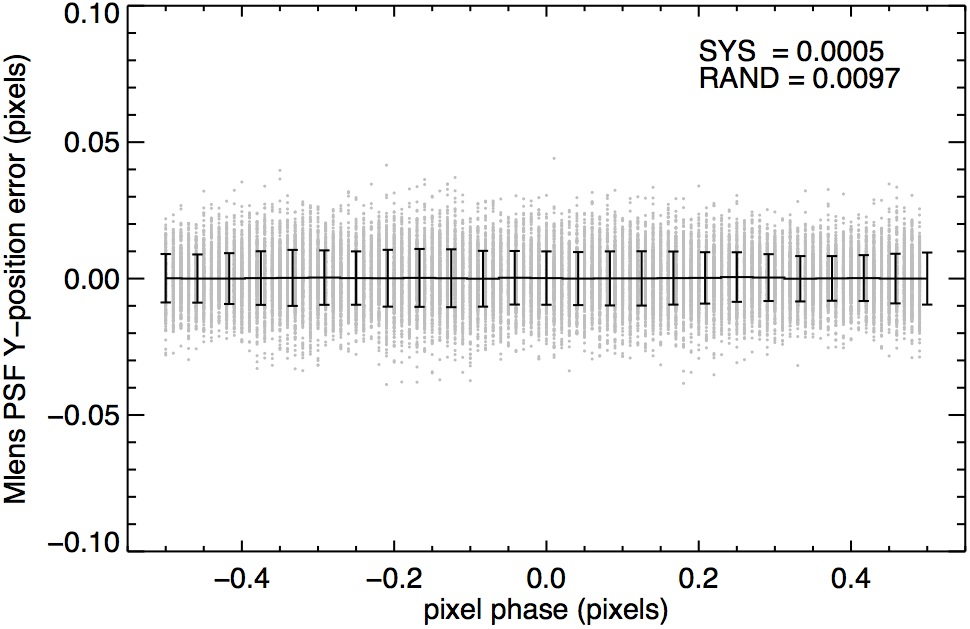} \\
\end{tabular}
\end{center}
\caption[example]{ \label{fig:gaussian_pixel_phase} \textbf{Top:} The X/Y (Left/Right) positioning systematic (SYS) and random (RAND) errors as a function of the PSF peak position when Gaussian fits to a simulated PSF are performed. \textbf{Bottom:} The errors when the high-resolution ePSF model is employed. Both the random and systematic errors are significantly decreased. This demonstrates the dangers of fitting undersampled PSFs. The grey dots indicate each positioning trial whereas the errorbars represent the 1$\sigma$ variation about the mean (stepped line), for a given bin. }
\end{figure}

An important application of the high-resolution ePSF models is their use in measuring the position of the under-sampled PSFs on the detector. Because we do not know the true form of the microlens PSF, we cannot simply evaluate the performance of our model. However, what we can demonstrate is the ability of our algorithms resolve a simulated detector-sampled PSF without pixel phase error, and how the centroids are incorrect when one assumes the PSF is Gaussian. To do this, we have taken a high-resolution ePSF model and evaluated it onto a detector sampling with the centroid at various positions over a pixel (a 50$\times$50 grid). Subsequently, both a 1\% pixel-to-pixel random error, signifying a pixel-to-pixel flat-fielding error, combined with a photon noise error, indicative of the photon noise estimated from our wavelength calibrations, was added to the image. The detector-sampled PSF was then fit using both an unconstrained Gaussian and our high-resolution lenslet model. Figure \ref{fig:gaussian_pixel_phase} shows the systematic and random errors encountered when fitting the PSF as a function of it's position (this is the pixel phase, which is the centroid of the PSF relative to the center of the pixel). Both systematic errors and a larger random error is observed when a Gaussian PSF is assumed, when compared to our model. As the photon noise is increases, the random error component also increases, but the systematic offset remains. Although the effects of systematic error are most apparent at shorter wavelengths, the effect is also seen in $K1$ band, indicating that centroiding using an high-resolution microlens ePSF model will always be more effective to remove the centroiding pixel phase errors.

\section{Applications of the high-resolution ePSFs}

The primary goals of developing the high-resolution ePSFs is to enable a greater understanding of the instrument and the systematic errors that are present in our datasets in order to increase the SNR of extracted data and reduce the effects of systematic error. To date, the most apparent systematic error is a Moir\'e pattern which is observed between adjacent spatial pixels (spaxels) in the datacubes (see Wolff et al this proceedings\cite{Wolffthis} for a figure). The suspected causes of this Moir\'e pattern are an imperfect wavelength solution and or an undesired effect of our current spectral extraction algorithm. 

The imperfection in the wavelength calibration determination is discussed in detail by Wolff et al (this proceedings), but is believed to be a result of two effects. The first is the centroiding accuracy of the individual peaks in the arclamp data, and the second is the assumption of a linear dispersion coefficient over the individual bandpasses. Not surprisingly, the two effects are related. The centroiding error introduced by fitting a Gaussian PSF on the individual peaks will result in different wavelength solutions between adjacent spectra. If one considers an extreme case of the spectrum consisting of a single emission peak, then this error would result in adjacent spectra assigning the peak emission to different wavelengths. Therefore when looking at a single wavelength in a reduced datacube, the wavelength calibration errors would appear as a Moir\'e pattern. Furthermore, if the dispersion is non-linear, as it is expected to be, this will also result in Moir\'e type pattern. As a result of the spectra being sampled differently over the field of view, we interpolate the data to a common set of wavelengths for each band. Because the extracted wavelengths will have some systematic error due to the assumption of a linear dispersion, this error will be manifested this interpolation. Both of these problems are solved with a higher centroiding accuracy, since it will enable a precise measurement of the non-linearity of the wavelength solution, and precise measurements of the peak for determining the spectral positions.

The second direct application of the microlens ePSF models is for developing a more detailed extraction algorithm. The current algorithm simply adds the intensities of a 3-pixel box centered on the peak of the microspectrum\cite{Jeromethis}. Depending on the position of the spectral peak, different amounts of flux for a given lenslet will fall in or out of this box, resulting in a Moir\'e type pattern in the single wavelength slices of the datacube. More sophisticated extraction algorithms are currently being developed based on a least squares method to disentangle the flux variations along the spectra and noise contributions (e.g. spectral crosstalk). The performance of this technique is dependent upon an accurate model of the lenslet PSF. More details can be found in the paper by Draper et al (this proceedings).

\subsection{Using the high-resolution ePSFs with the GPI Data Reduction Pipeline}

The next release of the GPI DRP will contain functions to work with the microlens PSFs.\footnote{We assume the reader is familiar with the basic operation of the GPI pipeline and key concepts such as reduction recipes and primitives. For description of these see \cite{perrinthis}} How the user wishes to interact with the model is dependent upon the task at hand, but the pipeline infrastructure already supports use of the high-resolution microlens PSFs for multiple applications. Routines to fit a detector sampled PSF using the high-resolution model, as is currently being performed when determining the wavelength solution already exist. Another routine  evaluates the high-resolution model into the detector space, as is performed for the spectral extraction routines currently under development. Details and instructions for using these functions will be included in the DRP documentation. The code to create the microlens PSFs is not part of the GPI DRP, however, the high-resolution model PSFs will be available to the public as part of the GPI calibration files.

One use of the microlens PSFs, which has not yet been developed, is the ability to put artificial signals into raw data cubes. Doing so will allow a direct measure of the systematic errors introduced by each step in the pipeline, and provide an excellent metric for pipeline performance.

\section{Implications for future instruments}

This demonstration deriving high-resolution ePSF models for lenslet-based integral field spectrographs shows how the effects of under-sampled PSFs of microspectra to be reduced. Under-sampling of the microspectra is often employed to enable a denser packing of the photons of the focal plane (e.g. more microspectra on a given detector). Instruments such as OSIRIS\cite{Larkin06a} on the Keck I telescope, employ a movable mask which allows a single microlens to be illuminated. By measuring the spectrum of each microlens individually, one can then use these in the extraction algorithm to compensate for crosstalk and flat field effects. OSIRIS sits on a Nasymth platform and is therefore not subject to flexure, therefore the calibration holds until the instrument is moved or cryo-cycled. For instruments that are required to undergo position changes, such as GPI, or any instrument \textit{not} on a Nasymth-type platform, including possible space-based lenslet IFS's such as is proposed for the WFIRST-AFTA mission\cite{Spergel13a}, such a calibration method is not feasible.

The work presented here demonstrates that if the high-resolution microlens ePSFs can be derived, a stringent flexure requirement after the lenslet array is not required so long as offsets to the wavelength solution can be derived. This is especially important when considering space-based missions that must be designed to survive the strong vibrations induced during launch and any thermal expansion/compression that may occur during during orbit or at different telescope pointing positions. Utilizing flexure to sub-sample the microlens PSFs may not be possible, however, if the microspectra are spaced such that each adjacent PSF is sampled differently (unlike GPI), the need for multiple flexure positions is reduced. Furthermore, the need for flexure can be entirely eliminated if the systematic offset of all lenslet PSFs is induced by a small change in wavelength since the lenslet PSFs will not vary significantly. For example, a 0.5 pixel displacement, the maximum required, is achievable on GPI with only a $\sim$20 nm change in wavelength. Using custom designed narrowband filters in conjunction with a white light source would make this achievable. Fewer filters could be used if comb filters were used in conjunction with the broadband filters. One could also conceive of using extended sources with sharp emission features, although finding sharp features in several band passes may prove challenging. Tunable lasers may also be an option depending upon the desired bandpass and required wavelength offset.


\section{Acknowledgments}
Our team mate Vincent Fesquet tragically passed away during the development of this work. Among many other contributions to GPI, his efforts were essential in getting the narrowband calibration data we used to derive the microlens PSF models. We would like to thank the staff of the Gemini Observatory for their assistance in the determination of the microlens PSFs. The Gemini Observatory is operated by the Association of Universities for Research in Astronomy, Inc., under a cooperative agreement with the NSF on behalf of the Gemini partnership: the National Science Foundation (United States), the National Research Council (Canada), CONICYT (Chile), the Australian Research Council (Australia), Minist\'erio da Ci\'encia, Tecnologia e Inova\c{c}\=ao (Brazil), and Ministerio de Ciencia, Tecnolog\'iae Innovaci\'on Productiva (Argentina).

\bibliography{main}   

\begin{thebibliography}{10}

\bibitem{Macintosh14a}
Macintosh, B., Graham, J.~R., Ingraham, P., Konopacky, Q., Marois, C., Perrin,
  M., Poyneer, L., Bauman, B., Barman, T., Burrows, A.~S., Cardwell, A.,
  Chilcote, J., De~Rosa, R.~J., Dillon, D., Doyon, R., Dunn, J., Erikson, D.,
  Fitzgerald, M.~P., Gavel, D., Goodsell, S., Hartung, M., Hibon, P., Kalas,
  P., Larkin, J., Maire, J., Marchis, F., Marley, M.~S., McBride, J.,
  Millar-Blanchaer, M., Morzinski, K., Norton, A., Oppenheimer, B.~R., Palmer,
  D., Patience, J., Pueyo, L., Rantakyro, F., Sadakuni, N., Saddlemyer, L.,
  Savransky, D., Serio, A., Soummer, R., Sivaramakrishnan, A., Song, I.,
  Thomas, S., Wallace, J.~K., Wiktorowicz, S., and Wolff, S., ``First light of
  the gemini planet imager,'' {\em Proceedings of the National Academy of
  Sciences}  (2014).

\bibitem{Macintoshthis}
Macintosh, B., Graham, J.~R., Ingraham, P., Konopacky, Q., Marois, C., Perrin,
  M., Poyneer, L., Bauman, B., Barman, T., Burrows, A.~S., Cardwell, A.,
  Chilcote, J., De~Rosa, R.~J., Dillon, D., Doyon, R., Dunn, J., Erikson, D.,
  Fitzgerald, M.~P., Gavel, D., Goodsell, S., Hartung, M., Hibon, P., Kalas,
  P., Larkin, J., Maire, J., Marchis, F., Marley, M.~S., McBride, J.,
  Millar-Blanchaer, M., Morzinski, K., Norton, A., Oppenheimer, B.~R., Palmer,
  D., Patience, J., Pueyo, L., Rantakyro, F., Sadakuni, N., Saddlemyer, L.,
  Savransky, D., Serio, A., Soummer, R., Sivaramakrishnan, A., Song, I.,
  Thomas, S., Wallace, J.~K., Wiktorowicz, S., and Wolff, S., ``The gemini
  planet imager: first light and commissioning,'' {\em Society of Photo-Optical
  Instrumentation Engineers (SPIE) Conference Series} {\bf 9148} (2014).

\bibitem{Chilcote12a}
{Chilcote}, J.~K., {Larkin}, J.~E., {Maire}, J., {Perrin}, M.~D., {Fitzgerald},
  M.~P., {Doyon}, R., {Thibault}, S., {Bauman}, B., {Macintosh}, B.~A.,
  {Graham}, J.~R., and {Saddlemyer}, L., ``{Performance of the integral field
  spectrograph for the Gemini Planet Imager},'' in [{\em Society of
  Photo-Optical Instrumentation Engineers (SPIE) Conference
  Series}{\nolinebreak\hspace{0.1em}]},  {\em Society of Photo-Optical
  Instrumentation Engineers (SPIE) Conference Series} {\bf 8446} (Sept. 2012).

\bibitem{Larkinthis}
Larkin, J.~E., Chilcote, J.~K., Aliado, T., Bauman, B.~J., Brims, G., Canfield,
  J.~M., Cardwell, A., Dillon, D., Doyon, R., Dunn, J., Fitzgerald, M.~P.,
  Graham, J.~R., Goodsell, S., Hartung, M., Hibon, P., Ingraham, P., Johnson,
  C.~A., Kress, E., Konopacky, Q.~M., Macintosh, B.~A., Magnone, K.~G., Maire,
  J., McLean, I.~S., Palmer, D., Perrin, M.~D., Quiroz, C., Rantakyrö, F.,
  Sadakuni, N., Saddlemyer, L., Serio, A., Thibault, S., Thomas, S.~J., Vallee,
  P., and Weiss, J.~L., ``{The Integral Field Spectrograph for the Gemini
  Planet Imager},'' {\em Society of Photo-Optical Instrumentation Engineers
  (SPIE) Conference Series} {\bf 9147} (2014).

\bibitem{Hardy08a}
{Hardy}, T., {Baril}, M.~R., {Pazder}, J., and {Stilburn}, J.~S.,
  ``{Intra-pixel response of infrared detector arrays for JWST},'' in [{\em
  Society of Photo-Optical Instrumentation Engineers (SPIE) Conference
  Series}{\nolinebreak\hspace{0.1em}]},  {\em Society of Photo-Optical
  Instrumentation Engineers (SPIE) Conference Series} {\bf 7021} (Aug. 2008).

\bibitem{Zimmerman11a}
{Zimmerman}, N., {Brenner}, D., {Oppenheimer}, B.~R., {Parry}, I.~R.,
  {Hinkley}, S., {Hunt}, S., and {Roberts}, R., ``{A Data-Cube Extraction
  Pipeline for a Coronagraphic Integral Field Spectrograph},'' {\em The
  Publications of the Astronomical Society of the Pacific}~{\bf 123},  746--763
  (June 2011).

\bibitem{Anderson00a}
{Anderson}, J. and {King}, I.~R., ``{Toward High-Precision Astrometry with
  WFPC2. I. Deriving an Accurate Point-Spread Function},'' {\em The
  Publications of the Astronomical Society of the Pacific}~{\bf 112},
  1360--1382 (Oct. 2000).

\bibitem{Marois06a}
{Marois}, C., {Lafreni{\`e}re}, D., {Doyon}, R., {Macintosh}, B., and {Nadeau},
  D., ``{Angular Differential Imaging: A Powerful High-Contrast Imaging
  Technique},'' {\em The Astrophysical Journal}~{\bf 641},  556--564 (Apr.
  2006).

\bibitem{Wolffthis}
Wolff, S.~G., Perrin, M., Maire, J., Ingraham, P.~J., Rantakyr\"o, F.~T., and
  Hibon, P., ``Gemini planet imager observational calibrations iv: Wavelength
  calibration and flexure correction for the integral field spectrograph,''
  {\em Society of Photo-Optical Instrumentation Engineers (SPIE) Conference
  Series} {\bf 9147} (2014).

\bibitem{Zackthis}
Draper, Z.~H., Marois, C., Wolff, S., Perrin, M., Ingraham, P., Ruffio, J.-B.,
  Rantakyr\"o, F.~T., and Goodsell, S.~J., ``Gemini planet imager observational
  calibrations ix: Least square inversion flux extraction,'' {\em Society of
  Photo-Optical Instrumentation Engineers (SPIE) Conference Series} {\bf 9147}
  (2014).

\bibitem{Jeromethis}
Maire, J., Ingraham, P.~J., Rosa, R. J.~D., Perrin, M.~D., Rajan, A.,
  Savransky, D., Wang, J.~J., Ruffio, J.-B., Wolff, S.~G., Chilcote, J.~K.,
  Doyon, R., Graham, J.~R., Greenbaum, A.~Z., Konopacky, Q.~M., Larkin, J.~E.,
  Marois, B. A. M.~C., Millar-Blanchaer, M., Patience, J., Pueyo, L.~A.,
  Sivaramakrishnan, A., Thomas, S.~J., and Weiss, J.~L., ``{Gemini Planet
  Imager Observational Calibrations VI: Photometric and Spectroscopic
  Calibration for the Integral Field Spectrograph},'' {\em Society of
  Photo-Optical Instrumentation Engineers (SPIE) Conference Series} {\bf 9147}
  (2014).

\bibitem{perrinthis}
Perrin, M., Maire, J., Ingraham, P.~J., Savransky, D., Millar-Blanchaer, M.,
  Wolff, S.~G., Ruffio, J.-B., Wang, J.~J., Draper, Z.~H., Sadakuni, N.,
  Marois, C., Fitzgerald, M.~P., Macintosh, B., Graham, J.~R., Doyon, R.,
  Larkin, J.~E., Chilcote, J.~K., Goodsell, S.~J., Palmer, D.~W., Labrie, K.,
  Beaulieau, M., Rosa, R. J.~D., Greenbaum, A.~Z., Hartung, M., Hibon, P.,
  Konopacky, Q.~M., Lafreniere, D., Lavigne, J.-F., Marchis, F., Patience, J.,
  Pueyo, L.~A., Soummer, R., Thomas, S.~J., Ward-Duong, K., and Wiktorowicz,
  S., ``Gemini planet imager observational calibrations i: overview of the gpi
  data reduction pipeline,'' {\em Society of Photo-Optical Instrumentation
  Engineers (SPIE) Conference Series} {\bf 9147} (2014).

\bibitem{Larkin06a}
Larkin, J., Barczys, M., Krabbe, A., Adkins, S., Aliado, T., Amico, P., Brims,
  G., Campbell, R., Canfield, J., Gasaway, T., Honey, A., Iserlohe, C.,
  Johnson, C., Kress, E., LaFreniere, D., Lyke, J., Magnone, K., Magnone, N.,
  McElwain, M., Moon, J., Quirrenbach, A., Skulason, G., Song, I., Spencer, M.,
  Weiss, J., and Wright, S., ``Osiris: a diffraction limited integral field
  spectrograph for keck,'' (2006).

\bibitem{Spergel13a}
{Spergel}, D., {Gehrels}, N., {Breckinridge}, J., {Donahue}, M., {Dressler},
  A., {Gaudi}, B.~S., {Greene}, T., {Guyon}, O., {Hirata}, C., {Kalirai}, J.,
  {Kasdin}, N.~J., {Moos}, W., {Perlmutter}, S., {Postman}, M., {Rauscher}, B.,
  {Rhodes}, J., {Wang}, Y., {Weinberg}, D., {Centrella}, J., {Traub}, W.,
  {Baltay}, C., {Colbert}, J., {Bennett}, D., {Kiessling}, A., {Macintosh}, B.,
  {Merten}, J., {Mortonson}, M., {Penny}, M., {Rozo}, E., {Savransky}, D.,
  {Stapelfeldt}, K., {Zu}, Y., {Baker}, C., {Cheng}, E., {Content}, D.,
  {Dooley}, J., {Foote}, M., {Goullioud}, R., {Grady}, K., {Jackson}, C.,
  {Kruk}, J., {Levine}, M., {Melton}, M., {Peddie}, C., {Ruffa}, J., and
  {Shaklan}, S., ``{Wide-Field InfraRed Survey Telescope-Astrophysics Focused
  Telescope Assets WFIRST-AFTA Final Report},'' {\em ArXiv e-prints}  (May
  2013).

\end{thebibliography}
\bibliographystyle{spiebib}   

\end{document}